\documentstyle[12pt]{article}
\author {Peter Huber \\Germanistisches Seminar\\
University of Heidelberg\\
Hauptstr. 207-209\\
D-69117 Heidelberg\\
Germany \\pethk@aol.com} 
\title {A Cosmologic Model Based on the Equivalence of
Expansion and Light Retardation \\ Part 1: Large-Scale Aspects} 
\begin{document}
\maketitle

\begin{abstract}
\noindent Proceeding from a homogeneous and isotropic Friedmann
universe a conceptional problem concerning light propagation in an
expanding universe is brought up. As a possible solution of this
problem it is suggested that light waves do not scale with $R(t)$.
With the aid of a Generalized Equivalence Principle a cosmologic
model with variable ``constants" $c$, $H$, and $G$ is constructed.
It is shown that with an appropriate variation of the Boltzmann
``constant" $k$ the thermal evolution of the universe is similar
to the standard model. It is further shown that this model explains
the cosmological redshift as well as certain problems of the standard
model (horizon, flatness, accelerated expansion of the universe).

\noindent PACS numbers: 98.80.Bp, 98.80.Hw, 04.20.Cv.

\noindent Keywords: cosmology, velocity of light, expansion
\end{abstract}

\section{Introduction}

There is no doubt that the standard model of cosmology is the
most successful approach in describing the universe as a whole
while accounting for numerous empirical data provided by
macro- and microphysical observations. Yet, the standard model
causes some complications, such as an initial singularity, a
flatness, horizon and density fluctuation problem, and some
more. Although some of them could be avoided by re-introducing
the cosmological constant $\Lambda$, which grants a variety of
models like inflationary, extended or hyperextended expansion,
the original simplicity and beauty of the model has disappeared.

In this paper it will be attempted to show that with the aid of
a simple yet physically productive principle of the same
category as the Equivalence or the Cosmological Principle it is
possible to obtain an elementary description of the universe
while avoiding the disadvantages of the standard model. The
validity of the Cosmological Principle is assumed. Thus the
scope is limited to Friedmann-Robertson-Lema\^\i tre-universes,
which are based on the principles of spatial isotropy and
homogeneity of the universe.

\section {A Problem of Light Propagation in Friedmann
universes}

Among the various problems of the standard Friedmann universe
there is one which had been mentioned occasionally in the past
but which had not been paid full attention so far. It concerns
the dependence of the dielectric and magnetic constants 
$\varepsilon_0$ and $\mu$ on  the gravitational potential
$\Phi$. The Maxwell equations in a constant gravitational field
imply the relation
\begin{equation}
\varepsilon = \mu = (1+2\Phi /c^2)^{-1/2} \label{0}
\end{equation}
(M{\o}ller [1969,1972], Landau/Lifshitz [1962,1975]).
In cosmology one is interested in large scale {\it eigen}
gravitation of the universe rather than in local gravitational
fields. It arises the question, in how far a universal $\Phi$
could vary. In Newtonian terms, the potential $\Phi_E$ of 
the proper gravitation of the universe  can be written as
\begin{equation}
\Phi_E = - M_u G / R \label{0a}
\end{equation}
where $M_u$ ist the universal mass, $G$ the gravitational 
constant and $R$ the extension (radius) of the universe.
In Friedmann models, $R$ is increasing as a function 
of time. This leads to a decreasing $\Phi_E(t)$, if one does
not postulate {\it ad hoc} assumptions between $M_u$,
$G$ and $R$ to keep the expression constant. According
to eq. (\ref{0}) we then obtain time-dependent ``constants"
$\varepsilon (t)$ and $\mu (t)$ with increasing values.
(A detailed mathematical analysis on the variation of
vacuum permittivity in Friedmann universes was given
by Sumner [1994].)
Because $(\varepsilon\mu)^{-1/2} = c$ the velocity of
light should also decrease. We should become familiar
with the idea, that a variable $c$ is not a heresy. Even
Einstein (1911) was ready to give up the absolute constancy
of light according to Special Relativity when he worked on
the influence of gravitation on light propagation. In this
paper he developed following equation which shows the
velocity of light as a function of the gravitational potential
$c = c_0 (1+\Phi / c^2)$. Years later he discovered, that
it is more comfortable to keep $c$ constant and to interprete
the elapsed time as a function of $\Phi$. This led to the
present form of General Relativity. However, $\Phi$ in
Einstein's equation was restricted to local gravitational
fields, which were regarded constant. Different values
of $\Phi$ result in different lapses of time. Friedmann's
solutions of the gravitaional equations in accordance with
Hubble's observations of cosmological red shifts imply,
as we have seen, a variable $\Phi_E(t)$. This is a
completely different situation. Just like Einstein in the
1910's we have the choice to interprete a globally 
decreasing $\Phi_E$ either as a global decrease of
the speed of light or in the sense of varying universal
time. Both alternatives, which base on mathematical
(i.e. ``absolute") scales, should be equivalent.
However, the mere idea that reference scales could
vary causes us to avoid the concept of absolute units
and to rely on physical scales as it is practically done
by defining space and time by electromagnetic (i.e.
physical) processes. That means in particular that
reference scales are not defined seperately but in
relation to each other. The following equation (next 
section, eq.[\ref{1}]) can be understood in this sense.

\section {A New Mechanism of Light Propagation and
its Relation to an Expanding Universe}

We define a relative decrease of the velocity of light 
with reference to an expanding universe. This implies
the following mechanism of light propagation. Let there
be a source $E$ emitting with a (mathematically) constant
frequency $\nu$. Since $c$ decreases continuously, the 
wave-length $\lambda$ decreases proportionally to $c$ at
emission. While traveling through the universe, however, 
$\lambda$ shall remain constant. Namely, it shall not expand
with $R(t)$ as in the standard model. The assumption of 
constant $c$ and non-expanding light waves
over cosmic distances means a retarded arrival of light in an
expanding universe. To illustrate that point: Let the distance
between an emission source $E$ and an observer $O$ at time $t_0$
be $m\lambda$ of a defined wave-length. At time $t_1 > t_0$ the
distance $EO$ has expanded with $R(t)$, however, not the single
$\lambda_r$. Therefore the distance at $t_1$ in terms of the
light wave will be $n\lambda_r$ (with $n>m$). The runtime of a
light beam with constant speed starting at $t_1$ will be longer
than light that left $E$ at $t_0$. This problem casts a new
light on the relation of expansion and the constancy of light:
In the standard model a huge ominous force is needed to maintain
expansion against gravity (vacuum energy, cosmological constant
and quintessence are synonyms of possible explanations). What,
if the nature of universal expansion were that in terms of
decreasing wave-lengths $\lambda$ at emission? Then there 
would be no more necessity for a repulsive force, which
maintains expansion. Usually length is defined by an
invariable rigid rod. This is, however, a mathematical definition, 
because {\it in praxi} on has to utilize a physical reference scale
like defined wave-lengths. There is no way for us to find
out whether the universe is ``really" expanding or whether 
the speed of light is ``really" decreasing. Both phenomena are
conditioning each other. This new principle of the equivalence 
of space expansion and light retardation can be formulated
\begin{equation}
Rc = const. \label{1}
\end{equation}
where $c$ is the seemingly retarded velocity of light in an
expanding universe and $R$ is from now on regarded as the
variable radius of the universe in the unit of meter (whereas 
the scale factor $a$ is dimensionless).

As the term ``equivalence principle" is applied to inert 
and heavy masses, the meaning of eq.(\ref{1}) will 
be referred to as ``Indiscernibility Principle" (IP). It has 
to be mentioned that the IP is not an additional assumption;
it just replaces $c=const.$ of standard cosmology. (By the way:
This concept avoids the paradox situation of a light beam from
a distant region of the universe entering a galaxy cluster: Its
waves should suddenly cease expansion and carry on with it 
when leaving the cluster.)

To demonstrate the mathematical aspects of the IR (eq.\ref{1}) we
regard light propagation in Robertson-Walker-Metrics (RWM). 
Because of homogeneity and isotropy of space we can regard a 
light trajectory with $\chi (t), \theta = const., \phi = const.$ which
reaches from $\chi(t_1) = 0$ to $\chi(t_0) = \chi$. Then the RWM
reduces to
\begin{equation}
ds^2 = c^2dt^2-R(t)^2d\chi^2 = 0 \iff d\chi = cdt/R(t)
\label{1a}
\end{equation}
We now regard two subsequent wave peaks. Both have to travel the
same distance from the source to the receiver (from $0$ to $\chi$):
\begin{equation}
\chi = \int_{t_1}^{t_0} {c(t)dt\over R(t)} =\int_{t_1+\delta
t_1}^{t_0+\delta t_0}{c(t)dt\over R(t)} \label{1b}
\end{equation}
The spatial distance of the two subsequent wave peaks is
$\lambda = c/\nu$. The temporal distance $\delta t$ relates to
the frequency $\nu$ via
\begin{equation}
\delta t = 1/\nu \label{1c}
\end{equation}
From eq.(\ref{1b}) follows
\begin{equation}
0 = \int_{t_0}^{t_0+\delta t_0}{c(t)dt\over R(t)} -
\int_{t_1}^{t_1+\delta t_1}{c(t)dt\over R(t)} = {c(t_0)\delta
t_0\over R(t_0)} - {c(t_1)\delta t_1\over R(t_1)} \label{1d}
\end{equation}
(Because $\delta \ll \chi$ we can regard $R(t)$ in the
integration interval as constant.) From (\ref{1c}) and
(\ref{1d}) we have for the emitted frequency $\nu_1$ and the
received frequency $\nu_0$
\begin{equation}
R(t_0)\nu_0 c(t_1) = R(t_1)\nu_1 c(t_0) \label{1e}
\end{equation}
The standard model assumes $c=const.$ in (\ref{1d}) and 
(\ref{1e}) and concludes $R\nu = const.$ We see, however, 
that ``stretching" waves are in general not an implication of 
the RWM but rather of the assumption of a constant $c$.

While in the standard model expansion and light propagation can
be imagined within the szenario of an expanding balloon, the 
suggested mechanism of light propagation corresponds to the 
conveyor belt mechanism where a pen swinging rectangular to 
the belt's movement, drawing waves upon it. The product of 
the pen's frequency and the wave-length gives the speed of the 
belt identified with $c$. A diminishing speed of the belt reduces 
the wave-lengths at emission or the other way round: reduced 
wave-lengths caused by the expansion discrepancy of $R$ and 
$\lambda$ at constant swinging frequency diminishes the 
product $c$. The particle version is similar: A photon is 
placed with constant frequency on the decelerating belt. 

This Retarded Light Model will in the following serve to treat
the cosmological questions. (Of course the balloon picture can
be further maintained if one is aware that in this model space 
expansion results in the fact that the coordinate system does 
not co-expand.)

\section {Determination of Time Variabilities of Cosmological
Quantities $H$, $c$ and $R$}

According to the Indiscernibility Principle (IP) it makes no
sense to propagate either a ``real" expanding universe or a ``real"
decreasing speed of light. The mathematical treatment combines
both properties: Eq.(\ref{1}) can be written $\dot R /R = -\dot c /c$.
With the use of $\dot R/R = H$, where $H$ represents the
Hubble parameter (rather than Hubble constant) we obtain the
expression
\begin{equation}
{\dot R\over R} = H = -{\dot c\over c} \label{2}
\end {equation}
We can write $R$ as a function of $c(t)$ and $H(t)$: $R=c/H$ and
obtain
\begin{equation}
Rc = c^2/H = const. \label{3}
\end{equation}
(In the following time-dependent quantities like $R(t)$, $c(t)$ are
abbreviated $R$, $c$, and so on. Certain value are characterized 
by indices like $R_0$.)

From (\ref{3}) the time dependence of $H$ can be calculated:
\begin{equation}
{d\over dt}\left ({c^2\over H}\right ) = 0 \label{4}
\end{equation}
and resolved to $\dot H$:
\begin{equation}
\dot H = -2H^2 \label{5}
\end{equation}

In regard of (\ref{2}) and (\ref{5}) the propagation of the
universe is
\begin{equation}
\dot R = c\label{6}
\end{equation}
This was formerly deduced by Milne (1948) in the frame of his
Kinematic Relativity. Differentiation of (\ref{6}) gives
according to (\ref{2})
\begin{equation}
\ddot R = \dot c = -Hc \label{7}
\end{equation}

From (\ref{5}) we can derive $H(t)$ by seperating of
variables and integration:
\begin{equation}
H(t) = {1\over 1/H_0 +2t}={H_0\over 1+2H_0 t}\label{7a}
\end{equation}

From (\ref{7}) and (\ref{7a}) the temporal variation of $c$
(in an expanding universe) is
\begin{equation}
c(t) = {c_0\over\sqrt {1+2H_0t}} \label{7b}
\end{equation}

Now we can specify the expansion of the universe. Its radius
$R$ has a time dependence of
\begin{equation}
R(t) = {c(t)\over H(t)} = {c_0\over H_0} \sqrt{1+2H_0t} =
R_0 \sqrt{1+2H_0t} \label{7c}
\end{equation}

\section{Cosmological Redshift in the Retarded Light Model}

Standard cosmology explains the cosmological redshift by 
expansion of $\lambda$ while the light beam travels a distance
$r$. In the Retarded Light Model (RLM) a certain wave-length
remains, once emitted, unchanged. However, $\lambda$ is a
function of time. The earlier it was emitted, the faster $c$ and
thus the larger $\lambda$ had been. When we observe a spectral
redshift, this is because the wave was emitted much earlier in
the history of the universe compared to the same electromagnetic
process observed at present on our planet. In the following a
simple deduction, based on conventional terms, is given.

Hubble's law, as it is commonly referred to, is 
\begin{equation}
v = Hr \label{8}
\end{equation}
Since the transmission speed of photons is finite we can
express any distance $r$ by the runtime of light.

\begin{equation}
r = ct \label{9}
\end{equation}
(Note that in standard cosmology $c$ is constant while in the
RLM it is a function of time here. As it is calculated with
functions and not with certain values, the expressions can be
left unintegrated at the present state.)

Eqs. (\ref{9}) and (\ref{8}) give
\begin{equation}
v = Hct \label{10}
\end{equation}
Applying (\ref{10}) to the expanding universe, $v$ means the
receding velocity of cosmic objects and $t$ the time their light
has taken to reach us. Division by $t$ gives the mean
acceleration $\bar a$ of cosmic objects in dependence of their
distance (standard interpretation) or, according to the RLM
equivalence principle eq. $(\ref{1}$), the deceleration of
$c$:
\begin{equation}
-v/t = - \bar a = \dot c = -Hc \label{11}
\end{equation}
Note that (\ref{11}) is deduced from standard terms only,
while
the identical result in (\ref{7}) uses a new concept of
expansion based on the IP.

In the standard model one can describe the cosmological redshift
in a first order approximation as a Doppler effect. This is also
possible within the RLM, as first shown in Huber (1992). In the
following we use the particle picture of the conveyor belt
model: We regard photons which are emitted by a radiation source
with a constant frequency $T_0^{-1}$ but with decreasing
velocity of light. The frequency of radiation shall be the same
everywhere in the universe and at all times. This condition
guarantees the spatial and temporal invariance of physical (or
chemical) processes and is just another formulation for the
Cosmological Principle stating the homogeneity of the universe.
The initial speed of the photon $n$ shall be $c_{0n}$. The index
$n$ means that the initial speed $c_0$ decreases with time. We
define a very first photon with the speed $c_{00}$. Then the
varying initial speeds can be written
\begin{equation}
c_{0n} = c_{00}-anT_0 \label{12}
\end{equation}
where $a$ is the (negative) acceleration. At time $t>nT_0$ the
$nth$ photon has the speed
\begin{equation}
c_n(t)=c_{0n}-a(t-nT_0)=c_{00}-at \label{13}
\end{equation}
At time $t$ photon $n$ has travelled a distance
\begin{equation}
r(t) = \int_{nT_0}^t c_n(t')dt'=c_{00}t-{a\over 2}t^2 -
c_{00}nT_0 + {a\over 2}(nT_0)^2 \label{14}
\end{equation}
This equation can be resolved for the time $t_r^{(n)}$ at which
photon $n$ has travelled the distance $r$:
\begin{eqnarray}
t_r^{(n)} & = & {1\over a}\left ( c_{00}-\sqrt{c_{00}^2 +
a^2(nT_0)^2-2ac_{00}nT_0-2ar}\right ) \nonumber\\
& = & {1\over a} \left ( c_{00}-(c_{00}-anT_0)
\sqrt{1-{2ar\over (c_{00}-anT_0)^2}}\right ) \label{15}
\end{eqnarray} 
where the positive root is excluded because one must have
$t_r^{(n)} = nT_0$ for $r=0$. The time interval $T_r^{(n)}$
between the arrival of two photons $n$ and $n+1$ at the
observer at distance $r$ is
\begin{equation}
T_r^{(n)}=t_r^{(n+1)}-t_r^{(n)} \label{16}
\end{equation}
Inserting (\ref{15}) in (\ref{16}) we get
\begin{equation}
T_r^{(n)}=-{1\over a}\left \{\begin{array}{c}
(c_{00}-a[n+1]T_0) \sqrt{1-{2ar\over (c_{00}- a[n+1]T_0)^2}}
\\
-(c_{00}-anT_0) \sqrt{1-{2ar\over
(c_{00}-anT_0)^2}}\hfill\end{array}\right \} \label{17}
\end{equation}
 
This time-distance-relation gives the absorbtion interval $T$ in
dependence on the distance $r$ from the radiation source and the
(absolute) time $nT_0$.

To obtain a Doppler interpretation we restrict (\ref{17}) to
relatively small distances $r$. (We know from standard
cosmology, that the Doppler interpretation of the cosmological
redshift is only valid for distances $r<0.5 R_0$.) Then we can
expand (\ref{17}) in powers of ${2ar\over (c_{00}-anT_0)^2} \ll 1$:
\begin{equation}
T_r^{(n)}\approx -{1\over a}\left \{-aT_0-ar\left ( {aT_0\over
c_{00}^2\left[1-{a(2n+1)T_0\over c_{00}}+ {a^2n(n+1)T_0^2\over
c_{00}^2}\right ]}\right ) \right \} \label{18}
\end{equation}
Under the additional assumption $c_{00}\gg anT_0$ this
expansion leads to
\begin{equation}
T_r^{(n)}\approx T_0\left( 1+{ar\over c_{00}^2}\right )
\label{19}
\end{equation}
or written with frequency $\nu_r = T_r^{-1}$:
\begin{equation}
\nu_r\approx {\nu_0\over 1+{ar\over c_{00}^2}} \label{20}
\end{equation}

The classical Doppler effect for the frequency of light
escaping from a source moving with velocity $v$ is
approximately given by
\begin{equation}
\nu = {\nu_0\over 1+{v\over c}} \label{21}
\end{equation} Interpreting the Hubble flow as Doppler
redshift we have to replace $v$ by $Hr$ and get:
\begin{equation}
\nu = {\nu_0\over 1+{Hr\over c}} \label{22}
\end{equation}
Comparing (\ref{22}) with (\ref{20}) we immediately find that
both expressions are equal for $c_{00} = c$ and $ar/c=Hr$ or 
\begin{equation}
a =  Hc \label{23}
\end{equation}
This calculation shows that within the
Retarded Light Model the cosmological redshift can be
interpreted as a Doppler effect just like in the standard
model. The acceleration parameter $a$ has been introduced
negatively in (\ref{12}) so it bears
no explicit sign. It can be associated with $-\dot c$. The
result of (\ref{23}) represents another independent
determination of the results of (\ref{7}) and (\ref{11}).

There is yet another method to obtain this result. In the
RWM we have for the distance D between an emitter and
an observer the following relation, restricted to the first
two powers of a Taylor expansion:
\begin{equation}
D = D(t_0) = R(t_0)\chi \simeq c(t_0-t_1) + {Hc\over 2}
(t_0-t_1)^2 \label{23a}
\end{equation}
where $\chi$ is the light trajectory from a remote
light source ($\chi=0$) to the observer ($\chi$). This 
expression is within its limits identical with the formula 
of accelerated movement $d=v_0t + at^2/2$. Thus
we can associate the acceleration $a$ with $Hc$.
Because the increase of distance corresponds to the
decrease of $c$ we have $a = -\dot c = Hc$. It seems
that there is hardly a chance to get around the conclusion
that the cosmological redshift is caused by a light
deceleration of $\dot c = -Hc$. The last three of the
four presented methods to determine $\dot c$ in order
to compensate for the cosmological red shift do not
make use of the IP, so this relation stated above can be
regarded as a consequence of the new interpretation
of the observed cosmological redshift.

\section{Conclusions from Friedmann's equations}

Varying constants $c$ and $g$ applied in General
Relativity usually lead to Brans-Dicke theories. Even
there the Friedmann solution holds under certain
preconditions, as was shown by Albrecht/Magueijo 
and Barrow. The situation in the RLM is different,
however. It deals with the fact that Friedmann's
equations as a solution of Einstein's gravitational
equations imply varying magnitudes like wave-lengths
and distances in the universe, while there still exist
non-varying scales (at least by definition). As a
consequence of the Friedmann scale variation the RLM
eliminates mathematically defined ``absolute" scales
and refers to physical processes only. So the RLM is not
a super theory replacing or changing General Relativity;
it is applicable within its Friedmann solution {\it only}
and it wouldn't make sense elsewhere. The RLM is
restricted to cosmological applications, where the
universe as a whole and its {\it eigen} gravitation
play a role. The description of local gravitational
effects (star or galaxy interactions, for example)
must follow the unaltered Einstein laws. Once one
is aware of the hierarchy General Relativity -- Friedmann 
solution -- RLM, it is obvious that the Friedmann
equations can be applied unaltered. There is just
one exception: the cosmological constant $\Lambda$,
which mediates in the standard model between gravitation
and expansion, is superfluous, since the RLM itself is
the theory of mediation. Friedmann's equations without 
$\Lambda$ are:
\begin{equation}
 \left ({\dot R \over R}\right )^2 = - {kc^2\over R^2} +
{8\pi\over 3}G\varrho \label{24}
\end{equation}
\begin{equation}
 2 {\ddot R \over R} = - \left ({\dot R \over R}\right )^2 -
{kc^2\over R^2} - 8\pi Gp \label{25}
\end{equation}

read as follows (with $R=c/H$, $\dot R = c$ and $\ddot
R=-Hc$):

\begin{equation}
H^2 = -kH^2+{8\pi\over 3}G\varrho \label{26}
\end{equation}

\begin{equation}
-2H^2 = -H^2-kH^2-8\pi Gp  \label{27}
\end{equation}

In the RLM the space factor $k$ depends on the relation between
the density $\varrho$ and the radiation pressure $p$, however,
not on the universal mass $M$. For $p=\varrho/3$ (radiation era)
$k$ must be $0$ to fit both (\ref{26}) and (\ref{27}).
Accordingly, for $p=0$ (matter dominated era) we have
necessarily $k=1$. This may indicate, that a change in space
structure must have happened during the evolution of the
universe. Hovever, because in the RLM the extension of the
universe $\dot R$ occurs with the velocity of light, the
photons, also moving with $c$, cannot maintain an internal
pressure of radiation. For this reason it is more likely that
the universe always had $p=0$ and $k=1$.

\subsection{A relation between the Hubble radius $R_H$ and the
Gravitational radius $R_G$:}
From (\ref{24}) follows in respect of $H=\dot R/R,\ \varrho :=
3M/4\pi R^3$ and $R_H := R=c/H$: 
\begin{equation}
{c\over H} (k+1) = {2GM\over c^2} \label{28}
\end{equation}
For $k=1$ and $R_G= GM/c^2$ this leads to
\begin{equation}
R_H = R_G \label{29}
\end{equation}

Within the RLM the identity of expansion radius $R_H$ (Hubble
radius) and gravitational radius $R_G$ is not a coincidence.
This will be pointed out in the next section.

\subsection{A relation between $H$ and $q$:}
Differentiation of $H=\dot R/R$:
\begin{equation}
\dot H = {\ddot RR - \dot R^2\over R^2} = {\ddot R\over R} - H^2
= H^2\left ({\ddot R R^2\over R\dot R^2}-1\right ) \label{30}
\end{equation}
With regard to $q=-\ddot RR/\dot R^2$ we have
\begin{equation}
\dot H = - H^2(q+1) \label{31}
\end{equation}

According to (\ref{5}) this equation is true for the
deceleration parameter
\begin{equation}
q=1 \label{32}
\end{equation}

\section{Gravitation, Universal Mass and ``Distant Masses"}

For the state $k=1$ equation (\ref{28}) can be solved to the
universal mass $M$:
\begin{equation}
M = {c^3\over GH} \label{33}
\end{equation}

Inserting $R$ for $c/H$ we obtain the so-called ``Mach principle"
\begin{equation}
{GM\over c^2 R} = 1 \label{34}
\end{equation}
as it was quantitatively formulated by Sciama (1959). This is
remarkable, because it shows, that within the Retarded Light
Model the Friedmann universe based on Einsteins theory of
gravitation is in full concordance with the Mach principle.  
Eq.(\ref{33}) leads to the energy equivalent of the universe
\begin{equation}
E = {c^5\over GH}\label{35}
\end{equation}
Equations (\ref{33}) and (\ref{35}) represent the sum of
condensed matter and radiation. From the principle of energy
conservation $\dot E= 0$ follows
\begin{equation}
{d\over dt}\left ({c^5\over GH}\right ) = 0 \label{36}
\end{equation}
With (\ref{5}) and (\ref{7}) we find
\begin{equation}
\dot G = -3GH \label{37}
\end{equation}
and
\begin{equation}
G(t) = G_0 (1+2H_0t)^{-3/2} \label{38}
\end{equation}
A varying gravitational ``constant" was assumed by Dirac (1937,
1938) propagating his and Eddington's (1946) ``large number
observations" ($10^{40}$-relations). (The first remarks on the
$10^{40}$-numbers were given 1923 by Weyl.) Gravitational
experiments, though, seem to have almost excluded the alteration
of $G$ stated above (Hellings et al., Damour et al.). One has to
consider, however, that the values of $G$ and $M$ cannot be
seperated in gravitational measurements (Canuto and Hsieh). Thus
we have, regarding eqs. (\ref{37}) and (\ref{40}) for the
product $MH$ a relative decrease of $-H$. If $c$ is needed for
determinations of $G$, we even have
\begin{equation}
{GM\over c} = {c^2\over H} = const. \label{39}
\end{equation}

With the deduced time variation of the universal ``constants"
$c$, $G$ and $H$ we can calculate the time dependence of the
universal mass $M_H$. According to (\ref{5}), (\ref{7}),
(\ref{33}) and (\ref{37}) (respectively to (\ref{36}) and $E=
Mc^2$) we get 
\begin{equation}
\dot M = 2MH = {2c^3\over G} = const. \label{40}
\end{equation}
and
\begin{equation}
M(t) = M_0(1+2H_0t) \label{41}
\end{equation}
An increase of universal mass has been proposed at first by
Dirac to explain a large number relation, later by Narlikar and
Arp to obtain a ``tired light" mechanism for a non-expanding
universe. They showed that when a nucleus increases in mass, the
wave-length of emitted photons decrease. This process occurs
similarly in the RLM.

We now define the {\it eigen} gravitation of the universe
$F_G$ as
\begin{equation}
F_G:= {M^2\over R^2} G \label{42}
\end{equation}
where $M^2$ represents the self attraction of the universal
matter at maximal distance, the universal radius $R$. With
(\ref{33}) and $R_H = c/H$ we obtain
\begin{equation}
F_G = {c^4\over G} = {8\pi\over\kappa} \label{43}
\end{equation}
where $\kappa$ is representing Einstein's gravitational
``constant" in the field equations of General Relativity. In a
next step we describe the entire energy of the universe $E$
exclusively as the work of its {\it eigen} gravitation.
\begin{equation}
E= F_G R \stackrel{(\ref{42})}{=} {M^2GH\over c} = Mc^2
\label{44}
\end{equation}
Solved to $M$, the last two terms give $M = c^3/GH$. This
expression has been deduced in a different way before. Here it
results from the question: How large must a mass $m$ be, that its
intrinsic energy is totally described by its proper gravitation? 

So far only the self attraction of the universe was concerned. How
does the universal ``background" gravitation affect a test mass $m$?
Inserting (\ref{33}) and $R=c/H$ in (\ref{0a}) we obtain the
gravitational potential
\begin{equation}
\Phi = -c^2 \label{44a}
\end{equation}
That means, that the entire energy of a test mass $m$ is determined
by the universal gravitational potential. In Special Relativity the
famous formula $E=mc^2$ was obtained by kinematic reflections.
Here it follows from the {\it eigen} gravitation of the universe. This
demonstrates full concordance of inertia and gravity. One  may
also call it ``identity", as will be shown with the following
considerations.

On the other hand we can define an acceleration force $F_a$ of
the universal mass caused by light retardation:
\begin{equation}
F_a := M\dot c = {c^3\over GH} (-Hc) = - F_G \label{45}
\end{equation}
This equation illustrates the principle of the equivalence of
ponderable and inertial mass. While Einstein presumed the
equivalence principle to proceed from Special to General
Relativity he did not provide an explanation. Such an
explanation is possible within the Retarded Light model: Let all
matter of the universe be located on the ${\cal
R}_3$-``surface" of a 4-dimensional space ${\cal R}_4$. As
stated above, light retardation is identical with isotropic
space expansion in ${\cal R}_4$, which causes an inertial
force on all mass particles on ${\cal R}_3$ towards the center
of ${\cal R}_4$. This inertial force is being registered as
gravitation in the ${\cal R}_3$-subsphere. This can be
illustrated with an analogous {\it gedanken} experiment
reduced by one dimension: Let us imagine an air balloon in
gravitation-free space, which is half-ways blown up, and let
us place a metal ball somewhere on its surface. We then blow up
the balloon quickly. Due to inertia the metal ball will be pressed
towards the balloon surface and cause a dent. Now we repeat the
experiment with two metal balls being located close to each
other on the surface. Then both balls will form a common dent in
which they start moving against each other just as in a
gravitational field. And indeed, a 2-dimensional observer on the
balloon surface ${\cal R}_2$ not noticing space expansion will
describe the phenomenon as gravitation, maybe even as
hypothetical but imperceptible space curvature just like
Einstein. Returning to the universal situation we can say that
space expansion in ${\cal R}_4$ causes inertia, which is being
percepted as gravitation in ${\cal R}_3$. Gravitation, on the
other hand, causes energy degradation of electromagnetical
processes, which can be understood as light retardation.
Finally, light retardation is via gravity identical with
expanding space. Because of that identity proposed by the RLM we
emphasize the point that {\it gravitation causes expansion},
while the standard theory expects the expansion to be {\it
delayed} by gravitation. 

A historical remark: In Newtonian physics the equivalence of
inert and ponderable mass was a pure coincidence. Einstein used
this fact as an Equivalence Principle. Its explanation in the
sense of their identity, however, is provided by the RLM.

In a further step we can write the inert mass $F_a$ as $M\ddot
R$ and obtain from (\ref{45}) the following classical
(Newtonian) differential equation for the equivalence
principle:
\begin{equation}
 M\ddot R + {M^2 G\over R^2} = 0 \label{47}
\end{equation}
Its formal solution $R(t)$ agrees with eq.(\ref{7c}). With the
values for $M$, $R$ and $\ddot R$ inserted, the identity of
inertia and gravity is obvious.

With the identity of $|F_a| = Hcm$ and $|F_G| = MmG/R^2
 = Hcm$ the energy $E_m$ of a test mass $m$ on the 
${\cal R}_4$-surface with the radius $R$ is the product
\begin{equation}
E_m = \int |F_a| dR = \int |F_G| dR = \int Hcm dR = mc^2
\label{47aa}
\end{equation}
This result completes the above considerations concerning
the interaction of the universal mass  with a test mass $m$.

These examples show the high degree of self-consistence of the
Retarded-Light-Universe. It reveals to be highly ``Machian". In
the course of its elaboration the equivalence of gravity and
inertia was generalized with the aid of the IP and further to
the equivalence of expansion and gravitation (resp. gravitation
and light retardation). Referring to all of these aspects we use
the term ``Generalized Equivalence Principle" (GEP).

\section{Friedmann Variables Versus Einstein Constants - An
Alternative Approach}

In the last section we have found $|\phi|=c^2$. Inserting this
result in M{\o}ller's expression for $\varepsilon$ and $\mu$ 
we obtain time-independent electromagnetical constants. 
With these values the velocity of light would also be 
constant. Is there a contradiction to the varying $c$ as 
found above?The answer could be: yes, but only insofar 
as one is willing to admit a contradiction between 
General Relativity and Friedmann's equations. With the
universal expansion time, with the cosmic background
radiation we have an absolute reference frame, which should
not exist in the Theory of Relativity. On the other hand one
can have the point of view, that quantities in the frame of GR
do not necessarily have the same meaning as in the frame of
a Friedmann universe. When $R$ varies in the Friedmann
model while it is constant in GR and others of its solutions,
why should not other quantities like $c$ and $G$ can be
constant in GR and {\it simultaneously} varying in the 
Friedmann universe? If one accepts this argument one can
ask how the light speed in the Friedmann model $c_F$
must vary so that Einsteins $c_E$ can be kept constant.
From equation (\ref{0}) follows
\begin{equation}
c_E = c_0 (1+2\Phi/ c^2)^{-1/2} \label{82}
\end{equation}
The gravitational potential is given by
\begin{equation}
\Phi = MG/R \label{83}
\end{equation}
We find that with increasing $R$, which is essential
in the Friedmann model, $c_E$ can be kept constant
in any case if $c_F^2$ varies proportionally to $\phi$.
We then obtain in the Friedmann units
\begin{equation}
\Phi = c^2 = {MG\over R}\times const. \label{84}
\end{equation}
Solved to $R$ we immediatly obtain the satisfying
result
\begin{equation}
R(t) = MG/c^2 \times const. = R_G \label{85}
\end{equation}
Only from adjusting $c_F$ in a way that $c_E$
remains absolutely constant, we achieve with
$const. = 1$ the identity of gravitational and
expansion radius! Replacing the universal mass
$M$ by its energy we have
\begin{equation}
EG/c^4 = R \label{86}
\end{equation}
This is, beside the factor $8\pi$, the contracted,
direction-independent form of Einsteins gravitational
equations. With $R=c/H$ we have $E=c^5/GH$ for
the total energy in the universe. Since this magnitude
remains constant we obtain the following differential
equation:
\begin{equation}
5{\dot c\over c} - {\dot H\over H} - {\dot G\over G}
= 0 \label{87}
\end{equation}
On the other hand we see from eq. (\ref{86}) that
$G/c^4$ must vary proportionally to $R$.
Differentiation of $G/c^4\propto R$ gives with
$GH\propto c^5$ the result
\begin{equation}
{\dot G\over G} - 4{\dot c\over c} \propto H \label{88}
\end{equation}
Inserting (\ref{87}) in (\ref{88}) yields
\begin{equation}
\dot R\propto c \label{89}
\end{equation}
or
\begin{equation}
R = c/H \propto \int cdt \label{90}
\end{equation}
Starting out from the absolute constancy of $c$
in a varying gravitational field we obtain the
same results as above. To achieve a complete
solution of the two differential equations
above, we still need the relative variation of 
one of the magnitudes $c_F$, $G_F$ or $H$.
With the considerations of sections 4 and 5
we adopt the relation $\dot c = -Hc$ from
observation (cosmological redshift) and have
the Retarded Light Model.

To avoid two values $c_E$ and $c_F$ one
could tentatively regard $c$ in (\ref{82}) as
constant. In this case $MG/R=MGH/c$ must 
be constant as well. The universal mass is
then constant, too, because of $E/c^2 = const.
=M$. With this we would obtain $G\propto R$
and $G\propto 1/H$. That yields for both an 
increasing universal radius $R$ and universal 
time $1/H$ the very unlikely case of an 
increasing gravitational ``constant" $G$.
Beside this, there would arise another problem.
Equation (\ref{82}) is recursive, since $c^{-2}$
can be replaced by $\varepsilon\mu$, and both
again by eq. (\ref{0}), and so forth. With the
distinction of $c_E$ and $c_F$ as suggested
above this recursion can be avoided.

The intention of this chapter was to demonstrate
that a constant $c_E$ in General Relativity implies
a variable $c_F$ in a Friedmann model with varying
$R$ respectively $\Phi$. We have adjusted $c_F$
in a way that $c_E$ remained constant. This led
to the solution of the horizon problem. Restricting
the varying constants $c$ and $G$ to the Friedmann
model we do not have to alter the Einstein equations
in the sense of Brans-Dicke-theories, since in this
frame these magnitudes remain constant, as shown.
The original approach was based on differing
mathematical and physical units in a Friedmann
universe. The alternative approach lined out in
this section does not scrutinize the problem of 
measurement but assumes different behaviour
of ``constants" in Einstein and Friedmann frame,
such as $c_E$ and $c_F$. Both approaches yield
the same results and are obviously identical.

\section{Emission of Light and Temporal Relations}

In a previous section we have described light propagation
in the universe in terms of the conveyor belt model. We
have assumed a light source with constant emission 
frequency $\nu$. This was in fact a mathematical
setting. As pointed out before, we do not rely on ``ideal"
scales like an {\it a priori} constant time. We rather
use a physical process like the electron's change from
one energetic level of  the atom to another. The question
is, whether the frequency changes in cosmic dimensions.
If so, this would affect the theoretical explanation of the
observed red shift.  In the Bohr model (we use the 
Hydrogen atom) the emitted frequency  $\nu$ from the 
$m$th to the $n$th level is given by
\begin{equation}
\nu = {m_e e^4\over 8\varepsilon^2h^3}\left ({1\over n^2}
- {1\over m^2}\right ) \label{47a}
\end{equation}
The Planck constant $h$ and the electric charge $e$ are
absolute constants in this frame (for example M{\o}ller
[1972], p. 416f). The dielectric parameter $\varepsilon$
has, as mentioned above, the temporal variation
\begin{equation}
\varepsilon(t) = \varepsilon_0\sqrt{1+2H_0t}\label{47b}
\end{equation}
(The magnetic field ``constant" $\mu$ has the same
time dependence. Both $\varepsilon$ and $\mu$ compute
the function of $c(t)$ as given in eq. [\ref{7b}].) It
remains the mass $m_e$ of the electron. If it varied 
proportional to the universal mass $M$ (eq. 
(\ref{41}), then the emission frequency of 
electromagnetic radiation would be constant, 
so that the mathematical time would be identical 
with the physical time. However, there is evidence
that the electronic mass varies slightly  (see part 2, 
section 5).

Regarding the conveyor belt model it is obvious  that
the observed frequency must decrease with runtime 
respectively distance from the source, since a remote
observer receives longer wave-lengths being emitted
in earlier times, while the speed of light is identic
everywhere in three-space ${\cal R}_3$. We determine
the variation of $\nu$ in dependence of the distance 
from the source. Inserting (\ref{1}) in (\ref{1e})
yields:
\begin{equation}
R(t) \nu(t)/c(t) = const. \label{47c}
\end{equation}
According to the values for $R(t)$ and $c(t)$ we 
have a variation of $\nu \propto 1/(1+2H_0t)$.
This result can be evaluated more precisely.
According to (\ref{1d}) we have
\begin{equation}
\int_{t_0}^{t_0+\delta t_0}{c(t)dt\over R(t)} -
\int_{t_1}^{t_1+\delta t_1}{c(t)dt\over R(t)} = 0 \label{47d}
\end{equation}
Integration yields
\begin{equation}
ln\left ({[1+2H_0(t_0+\delta t_0)](1+2H_0t)\over 
(1+2H_0t_0)[1+2H_0(t_1+\delta t_1)]}\right ) = 0 \label{47e}
\end{equation}
Because the counter must equal the denominator we
obtain the relation $(1+2H_0t_0)/\delta t_0 = (1+
2H_0t_1)/\delta t_1$ or, according to (\ref{1c}),
\begin{equation}
\nu_0(1+2H_0t_0) = \nu_1(1+2H_0t_1) \label{47f}
\end{equation}
Setting $\nu_0$ for $t_0=0$ and $\nu (t)$ for $\nu_1$
we obtain the expression
\begin{equation}
\nu (t) = {\nu_0\over 1+2H_0t} \label{47g}
\end{equation}
This equation corresponds to the situation of an
observer receiving light from sources of various
distances at the same time, as it is the case when
looking at the nightly sky. Comparing this result 
with eq. (\ref{7a}) we see that this delay in frequency 
as a function of runtime (or distance) represents the 
Hubble flow. Looking towards the past, cosmic 
time seems to elapse twice as fast as on earth. 
(This result may play a role for the determination 
of the Hubble parameter.) Accordingly, the age 
of the universe is $1/2H_0$. This seems to be  a 
very short time. It has to be remarked, however,
that this is a mathematical time with no physical
relevance. If we use a physical clock and define
the unit of time being equal to one electromagnetic
oscillation of a certain frequency $\nu$, then
we obtain for the number $N$ of oscillations, which
represent the physical time elapsed between
the present $t_0=0$ and $-1/2H_0$:
\begin{equation}
N = \int_{-1/2H_0}^0 {\nu_0dt\over 1+2H_0t}
= -\infty \label{47h}
\end{equation}
This physical time may not be ``equidistant"
in the mathematical sense, however, it provides
at least the comforting argument that the
universe has physically existed ``forever".

\section{Density of Matter and Radiation and its Temporal
Variation}

The density of matter $\varrho_m$ and the density of radiation
$\varrho_r=\varrho_m c^2$ are not distinguished in standard
cosmology because $c=const.$ is assumed. This leads to a
temporal variation of $\dot \varrho = (\varrho + P/c^2)(-3\dot
R)$ with the result that $\varrho_m(t) R(t)^3=const.$ and
$\varrho_r(t) R(t)^4=const.$ The decrease of $\varrho_r$ with a
power of 4 is explained by a redshift effect of wave-lengths in
addition to the three spatial dimensions. The RLM has no
expansion of wave-lengths. This fact must be clearly deduceable
from the field equations of GR. They are 
\begin{equation}
R_{\mu\nu} = -{8\pi G\over c^4} \left( T_{\mu\nu}-{T\over
2}g_{\mu\nu}\right ) \label{48}
\end{equation}
The matter distribution in the universe is described by the tensor
\begin{equation}
T^{\mu\nu} = \left ( \varrho + {P\over c^2} \right ) u^\mu
u^\nu - g^{\mu\nu}P \label{49}
\end{equation}
According to the Friedmann-Universe we have spatial homogeneity
of density $\varrho$ and pressure $P$:
$\varrho(r,t)=\varrho(t)$
and $P(r,t)=P(t)$. With these constraints the 00-component of
eq. (\ref{48}) is
\begin{equation}
3\ddot R = -{4\pi G\over c^4} (\varrho c^2 + 3P)R \label{50}
\end{equation}
The spatial components all lead to the equation
\begin{equation}
R\ddot R + 2\dot R^2 + 2kc = {4\pi G\over c^4}(\varrho
c^2-P)R^2 
\label{51}
\end{equation} 
For $P=\varrho c^2 /3$ both equations are identical only for
$k=0$. ($R=c/H$, (\ref{6}) and (\ref{7}) have been used here.)
This would result in a radiation density of 
\begin{equation}
\varrho_r = 3H^2 c^2/8\pi G =: \varrho_c\label{52}
\end{equation}
As mentioned above, because of (\ref{6}) it is doubtful, whether
a radiation pressure had existed in a radiation-dominated era. 
For $P=0$, a scenario without remarkable radiation pressure,
both equations are only identical for $k=1$. Here the radiation
equivalent of the matter density $\varrho_r$ is
\begin{equation}
\varrho_r = {3H^2 c^2\over 4\pi G} = 2\varrho_c \label{53}
\end{equation}
Equations (\ref{52}) and (\ref{53}) show an important result:
In both cases the temporal variation $\dot\varrho_r$ is
\begin{equation}
\dot\varrho_r = - 3H\varrho_r \label{54}
\end{equation}
Accordingly we have a temporal variation of matter density in
the case $P=0$ and of the matter equivalent of radiation
pressure in the case $P=\varrho c^2/3$ of
\begin{equation}
\dot\varrho_m = - H\varrho_m \label{55}
\end{equation}

The universal density can be derived from inserting (\ref{24})
in (\ref{25}) with $R=c/H$, (\ref{6}), and (\ref{7}). This
gives the so-called equation of state
\begin{equation}
\varrho c^2 + 3P = {3H^2c^2\over 4\pi G} \label{56}
\end{equation}
Assuming $P=0$ for the whole history of the universe we always
have the result of eq.(\ref{53}). And indeed, Weinberg (1972)
admits, that, ``if we give credence to the values $q_0 \simeq 1$
and $H_0 \simeq 75$ km/sec/Mps [\dots], then we must conclude
that the density of the universe is about $2\varrho_c$" (p.
476). In the RLM the critical density $\varrho_c$ is just a
definition and does not have the meaning of the standard model,
which has a strict opposition of gravitation and expansion (and
wonders why their values seem to equal each other so perfectly).

As the spatial volume $V$ of the universe is $M/\varrho$, we
obtain the result
\begin{equation}
V = {4\pi R^3\over 3} \label{57}
\end{equation}
This has been implicitly used for eq.(\ref{28}). In standard
cosmology this volume is a consequence of $k=0$, which
indicates
an equilibrium between gravity and expansion. This state of
equilibrium is an integral part of the RLM, in which the meaning
of $k$ differs from the standard model. Therefore the assumption
(\ref{57}) is subsequently justified. The next section will provide
further evidence that the Retarded Light-universe is not
``closed" as $k$ indicates.

\section{On the Acceleration of Expansion}

The discovery in the recent years that the luminosity of
supernovae is smaller than  their redshifts suggest
(Perlmutter et al., Riess et al.,), has brought much confusion
into cosmological research. Anything could be expected but an
accelerated expanding universe. On the contrary: in the standard
model expansion is expected to be delayed by the {\it eigen}
gravitation of the universe. The cosmological constant $\Lambda$
has been revived, models of antigravitation and even of a
ominous ``quintessence" (tracker field) have been suggested
(Turner, Wang et al., Picon et al, Caldwell et al., Ostriker
and Steinhardt). The RLM, however, predicts an acceleration.

First, it will be demonstrated that in the RLM a measured
redshift $z$ results in a larger distance $D$ as in the standard
model. This will explain the weaker luminosity observed at far
supernovae. Then a mathematical indication for an accelerating
universe will be given. As mentioned above, eq.(\ref{0}) is
valid in the standard model as well as in the RLM. This implies
\begin{equation}
R(t_1)\nu_1 = R(t_0)\nu_0 \label{65}
\end{equation}
where $\nu_1$ is the at time $t_1$ emitted and $\nu_0$ is the
received frequency at time $t_0$. In the standard model with
constant $c$ the frequency can be replaced by the wave-length
as follows: $R(t_1)\lambda_0 = R(t_0)\lambda_1$. The spectral
redshift $z$ is defined
\begin{equation}
z={\lambda_0-\lambda_1\over \lambda_1} = {\lambda_0 \over
\lambda_1} -1 = {R(t_0)\over R(t_1)} -1 \simeq {H\over
c}D_{Standard} + \dots \label{66}
\end{equation}
(The restriction to the linear term of distance $D$ does not
alter the result in the sense of argumentation.) The RLM has
at the point of emission:
\begin{equation}
\nu = {c(t)\over \lambda(t)} = const. \label{67}
\end{equation}
Insertion of (\ref{67}) in (\ref{65}) yields
\begin{equation}
{R(t_1)c_1\over\lambda_1} = {R(t_0)c_0\over\lambda_0}
\label{68}
\end{equation}
This leads to an altered determination of $z$ in the RLM:
\begin{equation}
z={R(t_0)\over R(t_1)} - 1 = {\lambda_0 c_1\over \lambda_1
c_0} - 1 = {H\over c_0} D_{RLM} + \dots \label{69}
\end{equation}
Since $c$ decreases, the speed of light at emission time is
always higher than at reception, thus we have $c_1 > c_0$.
Comparing (\ref{69}) with (\ref{66}) we have
\begin{equation}
D_{RLM} > D_{Standard} \label{70}
\end{equation}
In words: For the same measured redshift $z$ the RLM yields a
larger distance $D$ to cosmic objects like supernovae as the
standard model. This important result explains the observed
lack in luminosity.

The acceleration of expansion is kind of a natural process
within the RLM without need of any external forces. Because the
deceleration of $c$ the universal mass $M$ must increase for
$E_{univ} = Mc^2 = const.$ In (\ref{40}) and (\ref{41}) the
values of this process are given. Increasing mass of and in the
universe  means more gravitational delay of photons; they will
slow down even more. The growing deceleration rate, however, is
equivalent to a more and more accelerating universe, according
to the GEP. The latest determinations of redshift and luminosity
of supernovae can be understood as a confirmation of the
Generalized Equivalence Principle of expansion, light
retardation, and gravitation.

One more remark on extension, expansion, and accelerated
expansion: The extension of the universe is given by $\dot R =
c$ (see eq.\ref{6}). Its expansion is determined by the rate of
light deceleration $\ddot R = \dot c = -Hc$ (see eq.\ref{7}). We
obtain the acceleration rate of expansion by differentiation of
$\dot c$,
\begin{equation}
\ddot c = 3H^2c \label{71}
\end{equation}

\section{Problems of the Standard Model: Horizon, Flatness,
Density Fluctuation}

One of the classical problems of the standard model is the
incompatibility of the event radius (extension) with
expansion. The event radius ${\cal D}$ is given by ${\cal D}(t) =
ct$, constant $c$ provided. Thus the causal event horizon has a
time dependence of ${\cal D} = const.\times t$. The time dependence
of expansion, derived from Friedmann's equations is $R(t)=
const. \times t^{2/3}$. That means, that different parts of
the universe must have been less causally connected in the
past -- a contradiction to the demand of a homogeneous universe,
which produced just those equations found by Friedmann. 

In the RLM, both radii have the same temporal development. For
${\cal D}$ we have
\begin{eqnarray}
{\cal D}(t) & = & \int c(t)dt = \int {c_0 dt\over \sqrt{1+2H_0t}}
\nonumber\\
& = & {c_0\over H_0}\sqrt{1+2H_0t} = R_0 \sqrt{1+2H_0t} = R(t)
\label{72}
\end{eqnarray}
In particular we have from the origin of the universe 
($t=-1/2H_0$) up to the presence ($t=0$)
\begin{equation}
{\cal D}_0 = \int_{-1/H_0}^{0} {c_0 dt\over 
\sqrt{1+2H_0t}} = {c_0\over H_0}=R_0 \label{72b}
\end{equation}
The identity of both horizons in the RLM is impressively
confirmed; their time dependence is ${\cal D}(t) = R(t) = const.
\times t^{1/2}$. To verify this result we insert eq.(\ref{6}),
the functions of $c(t)$ (see eq.\ref{7b}), $G(t)$ (see
eq.\ref{38}) and 
\begin{equation}
\varrho_m(t) = {\varrho_{m0} \over\sqrt{1+2H_0t}} \label{72a}
\end{equation}
which follows from (\ref{55}), in Friedmann's first equation
(\ref{24}).
\begin{equation}
\dot R^2 = -kc^2 + {8\over 3}\pi G\varrho_m R^2 \label{73}
\end{equation}
Solved to $R$ and all of the above functions inserted yield
\begin{equation}
R = \sqrt{{3(1+k)c_0^2(1+2H_0t)\over 8\pi G_0\varrho_{m0}}} =
const. \times t^{1/2} \label{74}
\end{equation}
These calculations show that there is no horizon problem
within the frame of the RLM.

The Retarded Light Model also provides an explanation for the
present flatness of the universe. Dividing (\ref{24}) by $H^2$
and setting $\varrho_c = 3H^2/8\pi G$ we obtain the following
equation for the curvature of the universe:
\begin{equation}
{k\over R^2} = {H^2\over c^2}\times {\varrho (t) -
\varrho_c\over \varrho (t)} \label{75}
\end{equation}
($\varrho$ means the density of matter.) Multiplication of
both sides with $\varrho_c/\varrho$ yields:
\begin{equation}
{\varrho - \varrho_c\over \varrho} = {3kc^2\over 8\pi
G\varrho R(t)^2} = const. \label{76}
\end{equation}
This is valid for a matter dominated as well as for a radiation
universe. The constancy in (\ref{76}) is a consequence of the
known functions of time $c$, $G$, $\varrho$, and $R$. Thus we
get for the curvature after inserting the proportionality term
of (\ref{76}) into (\ref{75}):
\begin{equation}
{k\over R^2} \propto {H^2\over c^2} \times const. = R(t)^{-2}
\times const. \label{77}
\end{equation}
This result indicates that the curvature of the universe is
proportional to $R(t)^{-2}$. Possible fluctuations around
$\Omega = \varrho /\varrho_c = 1$ do not enlarge as in the
standard model but remain in the same relation. And another
remark: Since we know that in the RLM $k=1$, we can replace the
proportionality symbol by equality in (\ref{77}). Everything in
the RLM fits perfectly, whereas the standard model produces
sometimes -- as in this case -- weird results. To compensate the
missing (or very small) space curvature observed today the
standard model had to introduce the {\it ad hoc} process of an
inflationary universe, reviving the cosmological constant, which
causes even more trouble than it prevents (Abbott 1988).

Along these lines the remaining cosmological problems of the
standard model, such as the increasing density fluctuation
during expansion, can be solved.

\section{The Temperature of the Universe}

Since in the presented model electromagnetic waves do 
not scale with $R(t)$, the relation between the scale factor 
or radius of the universe and its temperature $T$ should 
differ from the standard model. However, here arises a 
problem, which concerns the standard model just as 
well as the RLM. The question is, from which spectral 
distribution the temperature of the universe shall be
determined. If the universe is regarded as a ``black box",
then the Planck distribution would be a good candidate. In an
expanding universe, however, the Planck distribution needs a
certain relation between $T$ and $R$ to maintain itself during
expansion. This condition is
\begin{equation}
T(t) \propto {1\over R(t)} \label{78}
\end{equation}
This sounds surprising because this relation is in the standard
model frequently derived from the Stefan-Boltzmann law
$\varrho_r = aT^4$ with $a=\pi^2k^4/15\hbar^3c^2$ and a
conclusion from one of the Friedmann equations $\varrho_r
\propto R^{-4}$. These two relations yield $T\propto 1/r$.
However, the use of the Stefan-Boltzmann law implies the
validity of the Planck distribution, whereas the Planck
distribution needs the relation (\ref{78}) to ``survive" in an
expanding universe. To break this vicious circle we must refer
to experimental determinations of the cosmic background
radiation (CBR), which show indeed a good approximation to a
Planck distribution around 2.73 $K$. Because of this
observational result the RLM will also rely on the relation
described in (\ref{78}), although the radiation density
$\varrho_r$ relates in a different way to $R$ (see
eq.\ref{54}) as in the standard model. The Stefan-Boltzmann
law solved to the universal temperature T(t) is
\begin{equation}
T(t) = \root 4 \of {{15 \hbar^3c^5\varrho_r\over \pi^2k^4}}
\label{79}
\end{equation}
where $\hbar$ is the Planck constant divided by $2\pi$, and $k$
is the Boltzmann constant. Most of the ``constants" under the
root symbol are functions of $t$. Since $R\propto \sqrt{t}$ the
temperature $T$ must be proportional to $t^{-1/2}$ to satisfy
(\ref{78}). The time dependence of $c$ and $\varrho_r$ are
given in (\ref{7b}) and (\ref{54}). The latter gives 
\begin{equation}
\varrho_r(t)= \varrho_{r0} (1+2H_0t)^{-3/2} \label{79a}
\end{equation}
after integration. If the Planck constant $h$ is assumed to be
constant, then the Boltzmann ``constant" $k$ must obey the
temporal variation
\begin{equation}
k(t) = {k_0\over \sqrt{1+2H_0t}}  \label{80}
\end{equation}
or differentiated 
\begin{equation}
\dot k = -kH \label{81}
\end{equation}
With the above conditions of $h$ and $k$ the relation (\ref{78})
between $T$ and $R$ holds, so that the thermal evolution of the
universe can be treated analogously to the standard model. It
has to be emphasized that the variation of the Boltzmann
``constant" $k$ is no additional assumption. It follows 
from the observation that the spectrum of CBR is distributed 
in a Planckian manner. As pointed out, the standard model
 also has to refer to that empirical fact, which is
problematic in a certain way, because the assumption of a
Planck distribution requires a radiation equilibrium, which 
should not exist in the standard model because of its horizon
problems. The suggested model, however, is Mach connected
and its radiation is much more in a state of equilibrium.

\section{Summary}

The introduction of a new mechanism of light propagation in 
a Friedmann universe. avoids the problems of the standard 
model and provides a unifying description of various
empirical facts. Its probably most important feature is that 
light waves do not scale with $R(t)$. The expanding universe 
is defined as expansion relative to a constant reference 
wave-length $\lambda_0$. Vice versa one could
speak of a deceleration of $c$ relative to an expanding 
Friedmann universe. Both views are treated as mathematically 
and physically indiscernible. This principle of equivalence of 
expansion and light retardation (IP) is expressed by 
$Rc=const.$ The relation of these two cosmological 
quantities allows a classification of all the various 
cosmological models into four categories:

(1) $c=const.$ and $R\ne const. \Rightarrow Rc\ne const.$:
These conditions represent the Big Bang cosmologies including the
standard model. 

(2) $c=const.$ and $R=const. \Rightarrow Rc = const.$: This
describes the classical Steady State models (Bondi, Gold, Hoyle).

(3) $c\ne const.$ and $R=const. \Rightarrow Rc \ne const.$:
These conditions are preferred by modern Steady State theories
including various Tired Light models.

(4) $c\ne const.$ and $R\ne const. \Rightarrow Rc=const.$:
This characterizes the Retarded Light Model (RLM).

Because the observed cosmological redshift is usually explained
by scaling light waves, it first had to be provided another
interpretation of the observed redshift. Its determined value
$\dot c=-Hc$ has been found to be in accordance with theoretical
considerations given before. Some important features of the RLM
were developed, including a general condition of a universe:
$M_{univ} = c^3/GH$. The universal density was found to be
$2\varrho_c$, where the critical density $\varrho_c$ is just a
definition taken from the standard model but otherwise
meaningless in the RLM, because it was shown, that expansion
and {\it eigen} gravitation of the universe correspond to each
other in a way that a universal mass increase would lead to a delay
in light propagation which is equivalent to accelerated expansion.
This mechanism is referred to as the Generalized Equivalence
Principle (GEP). It is obvious that there is no need of a
cosmological constant $\Lambda$ in the RLM.

It was shown that the RLM avoids a variety of problems of the
standard model, including the recently discovered mysterious
acceleration of expansion. The RLM,
based on General Relativity and the Cosmologic Principle of
homogeneity and isotropy, includes Sciama's version
of the Mach Principle. It combines various isolated arguments
which arose from dissatisfaction with some features of the
standard model (Dirac, Narlikar, Milne) and puts them in a
consistent frame. There is a good chance, that the Retarded
Light Model can even more.

\end{document}